\documentclass[prl,aps,preprintnumbers,amsmath,amssymb,showpacs,showkeys,twocolumn,floatfix]{revtex4}

\usepackage{graphicx}
\usepackage{dcolumn}
\usepackage{bm}
\usepackage{units}
\usepackage{revsymb}
\usepackage{natbib}
\usepackage{xspace}
\usepackage[american]{babel}

\begin{document}


\title{Bose-Einstein condensation of chromium}
\author{Axel Griesmaier}
\email{a.griesmaier@physik.uni-stuttgart.de}
\author{J\"org Werner}
\author{Sven Hensler}
\author{J\"urgen Stuhler}
\author{Tilman Pfau}

\affiliation{5. Physikalisches Institut, Universit\"at Stuttgart, 70550 Stuttgart, Germany.}
\homepage{http://www.physik.uni-stuttgart.de/institute/pi/5}

\date{\today}

\begin{abstract}
We report on the generation of a Bose-Einstein condensate in a gas of chromium
atoms, which will make studies of the effects of anisotropic long-range
interactions in degenerate quantum gases possible. The preparation of the
chromium condensate requires novel cooling strategies that are adapted to its
special electronic and magnetic properties. The final step to reach quantum
degeneracy is forced evaporative cooling of $^{52}$Cr atoms within a crossed
optical dipole trap. At a critical temperature of T$_c\approx$\unit[700]{nK},
we observe Bose-Einstein condensation by the appearance of a two-component
velocity distribution. Released from an anisotropic trap, the condensate
expands with an inversion of the aspect ratio. We observe critical behavior of
the condensate fraction as a function of temperature and more than 50,000
condensed {$^{52}$}Cr atoms.
\end{abstract}

\maketitle The essential properties of degenerate quantum gases depend on
range, strength and symmetry of the present interactions. Since the first
observation of Bose-Einstein condensation in weakly interacting atomic gases,
eight different elements have been Bose-Einstein
condensed~\cite{Anderson:1995a,Davis:1995,Bradley:1995,Fried:1998,Modugno:2001,Robert:2001,Weber:2003a,Takasu:2003}.
All these elements, mainly alkali atoms, interact dominantly via short-range
isotropic potentials. Based on this effective contact interaction, many
exciting phenomena have been studied~\cite{Weidemueller:2003, Anglin:2002}.
Examples are the realization of four-wave mixing with matter
waves~\cite{Deng:99a} as well as the observation of
vortices~\cite{Matthews:99a, Madison:00a} and solitons~\cite{Burger:99a,
Denschlag:00a, Khaykovich:2002a} in degenerate quantum gases. Bose-Einstein
condensates (BECs) with contact interaction have also been used to investigate
solid-state physics problems like the Mott-metal-insulator
transition~\cite{Greiner:2002a,Stoferle:2004}. Tuning the contact interaction,
the collapse and explosion ("Bosenova") of Bose-Einstein condensates has been
studied~\cite{Donley:2001a} and new types of quantum matter like a
Tonks-Girardeau gas have been
realized~\cite{Paredes:2004}.\\
In a chromium Bose-Einstein condensate, one can not only tune the short-range
contact interaction using one of the recently observed Feshbach
resonances~\cite{Werner2004} but also investigate effects of the long­range and
anisotropic dipole-dipole interaction. This becomes possible because, compared
to other Bose-condensed elements, the transition metal chromium has a unique
electronic structure. The valence shell of its ground state contains six
electrons with parallel spin alignment (electronic configuration: $[{\rm
Ar}]3d^54s^1$). For the bosonic chromium isotopes, which have no nuclear spin,
this gives rise to a total electronic spin quantum number of 3 and a very high
magnetic moment of 6~$\mu_{\rm B}$ ($\mu_{\rm B}$ is the Bohr magneton) in its
ground state ${^7{}S_3}$. Since the magnetic dipole-dipole interaction (MDDI)
scales with the square of the magnetic moment, it is a factor of 36 higher for
chromium than for alkali atoms. For this reason, dipole-dipole interactions
which have not yet been investigated experimentally in degenerate quantum gases
will become observable in chromium BEC. For example, it was shown
in~\cite{Giovanazzi:2003a} that the MDDI in chromium is strong enough to
manifest itself in a well pronounced modification of the condensate expansion
that depends on the orientation of the magnetic moments. Tuning the contact
interaction between $^{52}$Cr atoms close to zero will allow one to realize a
dipolar BEC~\cite{Baranov:2002b} in which the MDDI is the dominant interaction.
This way, many predicted dipole-dipole phenomena, like the occurrence of a
Maxon-Roton in the excitation spectrum of a dipolar BEC~\cite{Santos_Roton03}
or new kinds of quantum phase transitions~\cite{Goral2002a,Yi:2004} as well as
the stability and the ground state of dipolar
BEC's~\cite{ODell:2004a,Goral:2002,Santos:2000a} can be investigated
experimentally. Since also the MDDI is tuneable~\cite{Giovanazzi:2002a}, a
degenerate quantum gas with adjustable long- and short-range interactions can
be realized.\\
A chromium BEC is also unique with respect to technical applications of
degenerate quantum gases. As a standard mask material in lithographic
processes, chromium is a well suited element for atom
lithography~\cite{Oberthaler:03a}. It has already been used to grow
nanostructures on substrates by direct deposition of laser-focused thermal
atomic beams~\cite{McClelland:93a,Drodofsky:97a}. Furthermore, structured
doping has been demonstrated by simultaneously depositing a homogeneous matrix
material and laser-focused chromium~\cite{Schulze:01}. Performing the step from
incoherent thermal atomic beams to coherent atom sources (BEC's), promises to
increase the potential of atom lithography in a similar way like
the invention of the laser did in classical optical lithography.\\
Our preparation scheme combines magneto-optical, magnetic and optical trapping
techniques. It requires novel cooling strategies that are adapted to its
special electronic and magnetic properties and to the need to circumvent
relaxation processes originating from the dipolar character of the atoms. A
beam of chromium atoms is generated by a high temperature effusion cell at
\unit[1600]{$^o$C} and slowed down by a Zeeman slower. Using a continuous
loading scheme~\cite{Schmidt:2003a,Stuhler:2001} followed by an in-trap
Doppler-cooling stage~\cite{Schmidt:2003c}, we prepare ${\sim 1.3\times10^8}$
atoms in the energetically highest projection ${m_J=+3}$ of the ${^7S_3}$
ground state in a Ioffe-Pritchard trap. Subsequently, during \unit[13]{s} of
radio­frequency (RF) induced evaporation, the atoms are cooled in the magnetic
trap to a phase space density five orders of magnitude below from degeneracy.
The extraordinary large magnetic dipole moment of chromium leads to increasing
two-body loss in the form of dipolar relaxation\cite{Hensler:2003a} with
increasing spacial density of the cloud. This causes RF-evaporation to become
inefficient and prevents ${^{52}}$Cr from reaching the regime of quantum
degeneracy in a magnetic trap. To overcome this loss mechanism, the atoms have
to be transferred into the energetically lowest Zeeman substate ${m_J=-3}$. In
this state, energy conservation suppresses dipolar relaxation if the Zeeman
splitting is much larger than the thermal energy of the atoms. As chromium
atoms in states ${m_J<0}$ are repelled from regions with low magnetic fields,
these atoms can not be trapped magnetically. We therefore adiabatically
transfer the atoms into an optical dipole trap where the trapping forces are
independent of the Zeeman substate. The optical trapping potential is formed by
two beams produced by a \unit[20]{W} fibre laser at \unit[1064]{nm}. The
stronger horizontal trapping beam has a waist of \unit[30]{$\mu$m} and a power
of up to ${\sim}$\unit[9]{W}. The symmetry axis of the trapping potential
formed by this beam coincides with the axis of our magnetic trap to achieve the
largest overlap between the two trap volumes for efficient transfer into the
optical trap. Ramping up this beam to its maximum intensity during the final
step of the RF ramp, we are able to transfer ${1.8\times10^6}$ atoms into the optical trap.\\
\begin{figure}
\includegraphics[width=1\columnwidth]{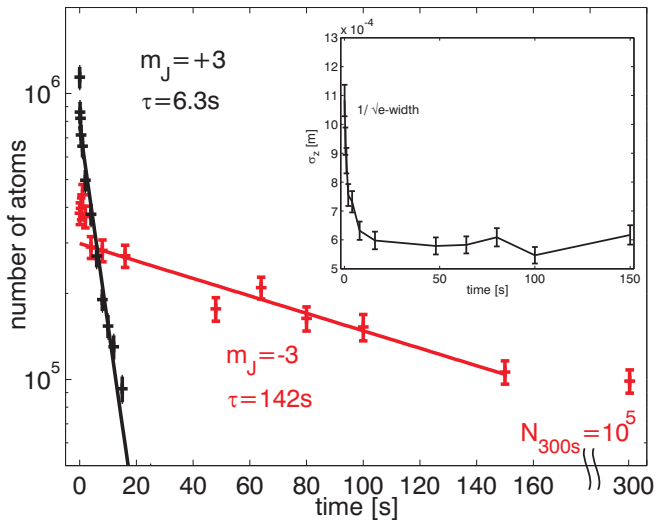}
\caption{ \label{fig:lifetime} Comparison of trap lifetimes before (black) and
after (red) pumping the atoms to the lowest Zeeman substate. The inset shows
the change of the axial size of the expanded cloud in time of flight after
different holding times.}
\end{figure}
The trap has a depth of \unit[130]{$\mu$K} and trap frequencies of
\unit[1450]{Hz} in radial and \unit[12]{Hz} in axial direction. After switching
off the magnetic trapping potential, we optically pump the atoms to the
${m_J=-3}$ state using a laser resonant to the ${^{7}S_3\rightarrow{}^{7}P_3}$
transition at an offset field of \unit[9]{G} in vertical direction. The
efficiency of the transfer is close to ${100\%}$ and is reflected in a dramatic
increase of the lifetime of the trapped gas from \unit[6]{s} in the ${m_J=+3}$
state to $>$\unit[140]{s} in the ${m_J=-3}$ state as shown in figure
\ref{fig:lifetime}. During all further steps of preparation, the offset field
is kept on in order to prevent thermal
redistribution among the other Zeeman states.\\
The optical transfer is followed by a \unit[5]{s} stage of plain evaporation
during which the number of atoms in the trap drops by 50${\%}$ and the phase
space density increases to ${\sim{}10^{-2}}$. This effect becomes visible in a
decrease of the axial size of the expanded cloud with increasing holding times
in the trap (see inset of figure \ref{fig:lifetime}). To increase the local
density and the elastic collision rate for evaporative cooling, a second beam
in vertical direction with a waist of \unit[50]{${\mu}$m} and a power of
$\sim$\unit[4.5]{W} is additionally ramped up adiabatically within the first
\unit[5]{s} in the optical trap. After the intensity of the horizontal beam is
reduced to ${70\%}$ of its initial value within \unit[10]{s}, about 300,000
atoms are trapped in this crossed trap. Forced evaporation towards the critical
temperature for the condensation proceeds now by gradually reducing the
intensity of the horizontal beam within \unit[11]{s}. Degeneracy is reached at
a remaining power of $\sim$\unit[800]{mW} in the horizontal beam. After holding
the atoms in the final trapping potential for \unit[25]{ms}, we switch off both
beams simultaneously and let the cloud expand freely for a variable time of
flight. The cloud is then detected using a standard absorption imaging
technique with a resonant probe beam propagating in the horizontal direction,
perpendicular to both trapping beams. For the imaging, the magnetic field is
rotated into the probe beam direction just
before releasing the atoms from the trap.\\
\begin{figure}
\includegraphics[width=0.9\columnwidth]{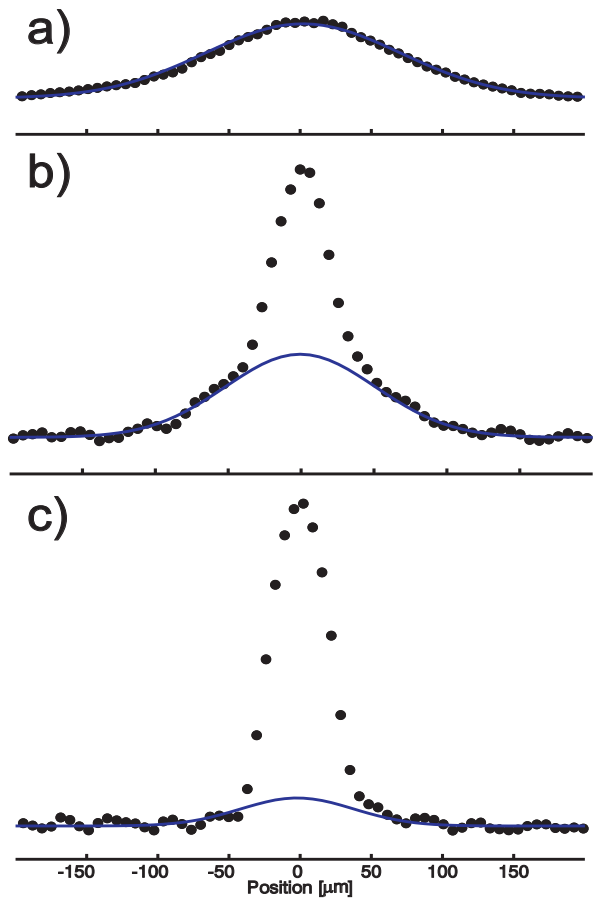}
\caption{ \label{fig:profiles} Density profiles from absorption images of atom
clouds taken after \unit[5]{ms} of ballistic expansion. A) thermal cloud at
\unit[1.1]{$\mu$K}, B) two-component distribution at \unit[625]{nK}, slightly
below $T_C$, C) nearly pure condensate with 47,000 atoms.}
\end{figure}
The onset of quantum degeneracy becomes visible in absorption images of the
relaxed chromium cloud. Figure \ref{fig:profiles} shows the profiles of the
cloud after 5ms of free expansion at different final powers of the horizontal
trapping beam before releasing the cloud. Figure \ref{fig:profiles}{\it(a)}
displays the situation at a final power of \unit[1.4]{W}. The cloud has the
Gaussian profile of a pure thermal distribution corresponding to a temperature
of $\sim$\unit[1.1]{$\mu$K}, very close to the critical temperature $T_C$. In
{\it (b)}, the cloud was released from a trap with a final power of
\unit[650]{mW} in the horizontal beam. The two­component distribution indicates
the presence of a Bose-Einstein condensate. The temperature of the remaining
thermal part of the cloud is \unit[625]{nK}. Reducing the laser power even
further to $\sim$\unit[370]{mW} leaves an almost pure condensate with more than
50,000 atoms and a non-Gaussian distribution as depicted in figure
\ref{fig:profiles}{\it(c)}. At the end of our evaporation ramp, the trap
frequencies in the visible axes are nearly equal. As expected, the expansion of
the condensate is almost isotropic and the aspect ratio of the expanded
condensate stays constant at $\sim$1.
\begin{figure}
\includegraphics[width=1\columnwidth]{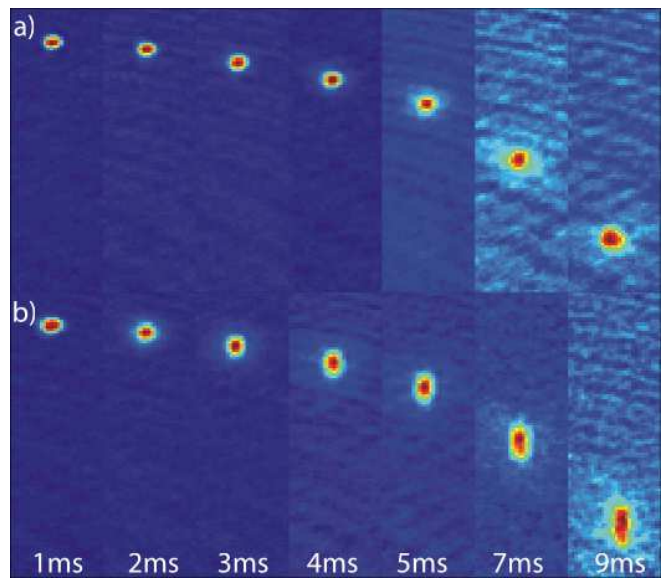}
\caption{ \label{fig:isoaniso} Time of flight series of absorption images with
expansion times from \unit[1]{ms} to \unit[9]{ms}. a) BEC released from an
almost isotropic trap, b) BEC released from an anisotropic trap.}
\end{figure}
A series of images of such a condensate taken after variable expansion times
between \unit[1]{ms} and \unit[9]{ms} is shown in figure
\ref{fig:isoaniso}{\it(a)}. The situation is different if the BEC is released
from an anisotropic trapping potential. The images in figure
\ref{fig:isoaniso}{\it(b)} show the expansion of the condensate for the same
time of flight series as before, but from a trap with a $\sim$20 times higher
intensity in the horizontal beam. In this case the BEC was prepared in the same
way as in {\it(a)} except that the trapping potential was deformed by
adiabatically increasing the horizontal laser power within \unit[250]{ms} after
the formation of the BEC. The series shows a clearly anisotropic expansion of
the cloud. From a cigar shaped trap in the horizontal direction, the aspect
ratio changes in time of flight to elongated in vertical direction. To show the
critical behaviour at the point of emerging degeneracy, we determine the number
of atoms in the condensate and the thermal fraction separately. We extract
these numbers by fitting a two-component distribution function to the density
profiles of the clouds at different final laser powers.
\begin{figure}
\includegraphics[width=1\columnwidth]{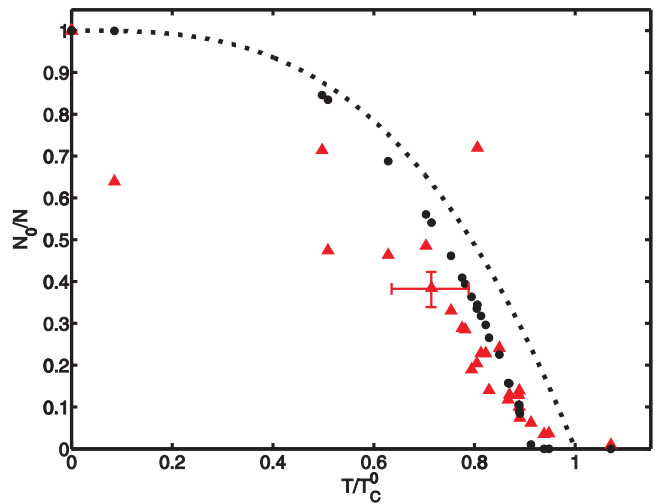}
\caption{ \label{fig:ttc} Condensate fraction ($N_0/N$) dependence on
temperature relative to the transition temperature of an ideal gas
($T/T_C^{0}$) given by equation \ref{eqn:NN0}. Red triangles represent our
measured data. Black circles represent the predicted fraction following
equation \ref{eqn:TCiafin} including corrections due to finite size effects and
the contact interaction The dashed curve shows the dependence for the ideal
gas. Error bars are mainly due to uncertainties in the measurement of the trap
frequencies, temperature and number of atoms.}
\end{figure}
In figure \ref{fig:ttc}, the fraction of condensed atoms in the total number of
atoms $(N_0/N)$ is plotted versus the ratio of the temperature of the remaining
thermal part to the critical temperature $(T/T_C)$. When we approach the
critical temperature from above $(T/T_C>1)$, the kink in the condensate
fraction plot marks the onset of Bose-Einstein condensation and provides an
experimental value for the critical temperature of $T_{exp}\sim$\unit[700]{nK}.
Based on the trap frequencies, number of atoms and temperature, we have also
calculated the expected condensate fraction for an ideal gas following the
equation
\begin{equation}
\label{eqn:NN0} {\frac{N_0}{N}=1-{\left(\frac{T}{T_C}\right)}^{3}}
\end{equation}
with
\begin{equation}
{T_C^{0}\approx0.94\frac{\hbar\omega}{k_B}N^{1/3}}
\end{equation}
being the critical temperature. When finite size effects as well as a
correction arising from the contact interaction~\cite{Giorgini:96} are taken
into account, the critical temperature is shifted to lower temperatures:
\begin{equation}
\label{eqn:TCiafin}
T_C=T_C^0+\delta T_C^{int}+\delta T_C^{fs}
\end{equation}
where ${\delta T_C^{fs}=-0.73\frac{\overline{\omega}}{\omega}N^{-1{}/3}T_C^0}$
is a shift in the critical temperature due to the finite number of atoms and
${\delta T_C^{int}=-1.33\frac{a}{a_{HO}}N^{1{}/6}T_C^{0}}$ takes into account
the contact interaction. Here $a=105a_0$ is the chromium scattering
length~\cite{Werner2004}, $a_0$ being Bohr's radius, $a_{HO}$ is the harmonic
oscillator length, $T$ is the temperature of the thermal cloud, ${\omega}$ is
the geometric and ${\overline{\omega}}$ the arithmetic mean of the trap
frequencies. These expected values are represented by black dots in figure
\ref{fig:ttc} and demonstrate a good agreement of our data with the
predicted dependence.\\
In conclusion, we have demonstrated Bose-Einstein condensation of chromium
atoms. We produce condensates with more than 50,000 chromium atoms, which is a
very good basis for a series of promising experiments on the dipolar character
of this novel quantum gas. We expect that both, long and short range
interaction, can be tuned by magnetic fields. As chromium is a standard
material in atom lithography, we also expect that an atom laser of chromium
will have applications in lithography, possibly even enabling controlled
deposition of single atoms.

\begin{acknowledgments}
We thank all members of our atom optics group for their encouragement and
practical help. We acknowledge especially the contributions of Piet Schmidt and
Axel G\"orlitz at earlier stages of the experiment. We thank Luis Santos, Paolo
Pedri, Stefano Giovanazzi and Andrea Simoni for stimulating discussions. This
work was supported by the SPP1116 of the German Science Foundation (DFG).
\end{acknowledgments}

\bibliography{./crbib}
\end{document}